\documentclass[]{interact}

\usepackage{epstopdf}% To incorporate .pdf illustrations using PDFLaTeX, etc.
\usepackage[caption=false]{subfig}% Support for small, `sub' figures and tables
\usepackage{amsmath, amsfonts, amssymb}
\usepackage{multirow}
\usepackage{threeparttable}
\usepackage{color}

\usepackage{hyperref}
\usepackage[numbers,sort&compress]{natbib}% Citation support using natbib.sty
\bibpunct[, ]{[}{]}{,}{n}{,}{,}% Citation support using natbib.sty
\renewcommand\bibfont{\fontsize{10}{12}\selectfont}% Bibliography support using natbib.sty
\makeatletter% @ becomes a letter
\def\NAT@def@citea{\def\@citea{\NAT@separator}}% Suppress spaces between citations using natbib.sty
\makeatother% @ becomes a symbol again

\theoremstyle{plain}% Theorem-like structures provided by amsthm.sty

\theoremstyle{definition}

\theoremstyle{remark}

\begin{document}

\articletype{Regular article}% Specify the article type or omit as appropriate

% \title{A Bayesian search for a formula describing hydroxide ion conductivity in anion-conductive polymer membranes}
\title{
Bayesian sparse modeling for interpretable prediction of hydroxide ion conductivity in anion-conductive polymer membranes
}

% Ryo Murakami and Kenji Miyatake and Ahmed Mohamed Ahmed Mahmoud and Hideki Yoshikawa and Kenji Nagata
\author{
    \name{
        Ryo Murakami\textsuperscript{a}\thanks{CONTACT Ryo Murakami. Email: MURAKAMI.Ryo@nims.go.jp},
        Kenji Miyatake\textsuperscript{b,c}\thanks{CONTACT Kenji Miyatake. Email: miyatake@yamanashi.ac.jp},
        Ahmed Mohamed Ahmed Mahmoud\textsuperscript{b,d},
        Hideki Yoshikawa\textsuperscript{a},
        Kenji Nagata\textsuperscript{e}
    }
    \affil{
        \textsuperscript{a}Research Network and Facility Services Division, National Institute for Materials Science, Tsukuba 305-0044, Ibaraki, Japan;\\
        \textsuperscript{b}Clean Energy Research Center, University of Yamanashi, Kofu 400-8510, Yamanashi, Japan;\\
        \textsuperscript{c}Department of Applied Chemistry, Waseda University, Tokyo 169-8555, Japan;\\
        \textsuperscript{d}Office of Institutional Advancement and Communications, Kyoto University, Kyoto 6068501, Japan;\\
        \textsuperscript{e}Center for Basic Research on Materials, National Institute for Materials Science, Tsukuba 305-0044, Ibaraki, Japan
    }
}

\maketitle

\begin{abstract}
Anion-conductive polymer membranes have attracted considerable attention as solid electrolytes for alkaline fuel cells and electrolysis cells. Their hydroxide ion conductivity varies depending on factors such as the type and distribution of quaternary ammonium groups, as well as the structure and connectivity of hydrophilic and hydrophobic domains. In particular, the size and connectivity of hydrophilic domains significantly influence the mobility of hydroxide ions; however, this relationship has remained largely qualitative. In this study, we calculated the number of key constituent elements in the hydrophilic and hydrophobic units based on the copolymer composition, and investigated their relationship with hydroxide ion conductivity by using Bayesian sparse modeling. As a result, we successfully identified composition-derived features that are critical for accurately predicting hydroxide ion conductivity.
% アニオン導電性高分子膜は、アルカリ型燃料電池や電解セルの固体電解質として注目されている。導電率は、四級アンモニウム基の種類や配置、親水・疎水部の構造とその連結性により変化する。特に、親水性ドメインの大きさや連結性は水酸化物イオンの移動に影響するが、その関係は定性的理解にとどまっていた。そこで本研究では、親水部・疎水部を構成する主要元素の数を共重合比から算出し、特徴量として導電率との関係を解析した。その結果、導電率予測に重要な組成特徴量を示すことができた。
\end{abstract}

\begin{keywords}
anion-conductive polymer membranes; Materials informatics; Data-driven science; Sparse modeling; Bayesian inference
\end{keywords}

\section{Introduction}
Anion-conductive polymer membranes are promising candidates for use as solid electrolytes in alkaline energy devices, such as fuel cells and water electrolysis cells. In particular, anion exchange membrane water electrolysis systems, which can produce green hydrogen efficiently by utilizing renewable energy sources, are being actively investigated worldwide as a core technology for realizing a carbon-neutral hydrogen society. For such applications, desirable properties of anion-conductive polymers include anion conductivity comparable to that of alkaline aqueous electrolytes, the ability to form thin membranes (thickness $< 50 \mu m$) with sufficient mechanical strength, gas barrier properties against hydrogen and oxygen, and water permeability and alkaline stability under low-humidity conditions. Among these, hydroxide ion conductivity is the most critical property, as it directly determines the voltage efficiency of anion exchange membrane water electrolysis systems.

Anion-conductive polymers are generally composed of hydrophobic polymer backbones functionalized with cationic groups. While a variety of backbones—including aliphatic and aromatic types—are available, it is known that structures without hetero-bonds such as ether linkages in the main chain are preferable from the viewpoint of alkaline stability \cite{Park2024}. Onium salts are commonly employed as cationic functional groups due to their synthetic accessibility, ease of incorporation into polymers, and adequate hydroxide ion dissociation; among them, quaternary ammonium groups are known to provide high anion conductivity. In general, the ion exchange capacity (IEC) of quaternary ammonium groups has a major impact on conductivity—higher IEC values typically correspond to higher hydroxide ion conductivity in thin membranes. However, even at similar IEC values, the conductivity can vary depending on the polymer backbone, the type of ammonium group, and their combination, as well as the connectivity of hydrophilic and hydrophobic domains. This is because quaternary ammonium groups and water molecules can aggregate to form hydrophilic domains, and the size, connectivity, and tortuosity of these domains significantly affect hydroxide ion mobility.

Hydrophilic domain structures can be analyzed by various experimental techniques. For instance, small-angle X-ray scattering (SAXS) provides information on the shape and periodicity of hydrophilic domains, while transmission electron microscopy (TEM) allows direct observation of domain size and connectivity \cite{Sriram2023}. Atomic force microscopy (AFM) can distinguish hydrophilic and hydrophobic domains on the membrane surface based on phase contrast, and electrochemical AFM can visualize ion-conductive pathways as current responses during oxidation or reduction reactions. Although these techniques provide valuable insights into the phase-separated morphology of the membrane, most discussions to date have remained qualitative, and a quantitative understanding has not been established. Moreover, cases where membranes exhibit similar morphologies and IECs but differ in hydroxide ion conductivity remain unresolved.

\begin{figure}
    \centering
    \includegraphics[width=\linewidth]{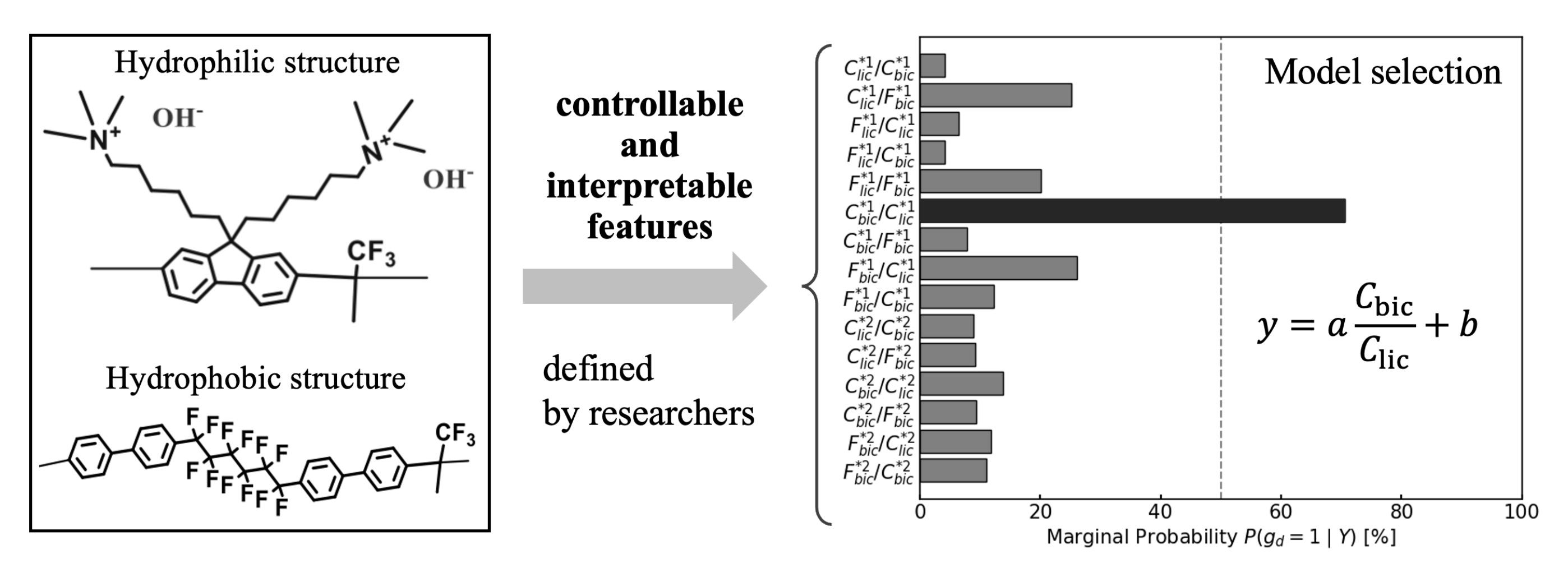}
    \caption{ Overview diagram of this study. We design controllable and interpretable candidate features from the chemical structures shown on the left side. Then, as shown on the right side, we explore equations that describe material properties (conductivity) by using model selection based on the marginal log-likelihood in Bayesian estimation framework. }
    \label{fig:abstract}
\end{figure}

In this study, we aimed to quantitatively understand the variation in conductivity among a series of anion-conductive polymer membranes previously developed by our group \cite{MiyatakeData2018, MiyatakeData2022, MiyatakeData2023, MiyatakeData2024}. We calculated the number of key constituent elements in the hydrophilic and hydrophobic units, taking into account the copolymer composition, and examined their potential as predictive features for hydroxide ion conductivity. Our goal was to construct an interpretable and low-dimensional predictive model using a linear analytical framework. Since the set of important features is not known a priori, we generated multiple candidate features and performed both conductivity prediction and feature selection simultaneously, as illustrated in Figure \ref{fig:abstract}.

% アニオン導電性高分子膜はアルカリ型エネルギーデバイス用の固体電解質として用いることが可能で、燃料電池や電解セルなどへの応用が期待されている。特に、アニオン膜型水電解セルは再生可能エネルギーを電源として用いることにより高効率にグリーン水素を製造することができ、カーボンニュートラルな水素社会の基幹技術と成り得ることから、世界中で活発に研究が進められている。これらの用途におけるアニオン導電性高分子に必要とされる物性は、アルカリ水溶液と同程度のアニオン導電率、50μｍ以下の膜厚での製膜性と機械強度、生成する水素や酸素のバリア性、低含水率状態における水透過性とアルカリ安定性、などである。特に、アニオン導電率はアニオン膜型水電解セルの電圧効率を決める因子であり、最も重要な物性である。

% アニオン導電性高分子膜は、一般的に疎水性高分子にカチオン性官能基を置換した構造から成り立っている。疎水性高分子としては脂肪族系や芳香族系などから多数の選択肢があるが、アルカリ安定性の観点からエーテル結合などのヘテロ結合を主鎖に含まない構造が好ましいことが知られている。カチオン性置換基は、合成方法の容易さ、高分子への置換のしやすさ、水酸化物イオンの解離度、の観点からオニウム塩がよく用いられており、特に、四級アンモニウム基が高いアニオン導電率を示す。一般的に、四級アンモニウム基のイオン交換容量（IEC）はアニオン導電率に最も大きな影響があり、IECが高くなると、その薄膜のアニオン導電率は高くなる。IECが同じであっても、疎水性高分子主鎖、四級アンモニウム基の種類とそれらの組み合わせ、親水部単位構造や疎水部単位構造の連結性の違いによって、アニオン導電率が異なることがある。薄膜中で四級アンモニウム基および水分子が凝集して親水部ドメインが形成されるが、この親水部ドメインの大きさや連結性・ねじれが水酸化物イオンの移動度に影響するからである。

% 親水部ドメイン構造は様々な実験手法によって解析することができる。例えば、小角X線散乱では親水部ドメインの形状や周期性を知ることができ、透過型電子顕微鏡では親水部の大きさと連結性を直接観測することができる。原子間力顕微鏡では位相角の違いから薄膜表面の親水部ドメインと疎水部ドメインの領域を識別することができ、電気化学AFMでは水酸化物イオンが導電する領域を酸化または還元反応の電流値として可視化することができる。これらの手法によって得られた親水部ドメインと疎水部ドメインから成る相分離構造に関する情報は、各種アニオン導電性高分子膜のアニオン導電率の違いを理解するのに非常に有効であるが、多くの議論は定性的であり、定量的な理解は得られていない。また、IECが同程度で同様なモルフォロジーを示すにもかかわらず、アニオン導電率が異なるケースも多く、その理由は明確ではない。

% 本研究では，これまで我々が開発してきた一連のアニオン導電性高分子薄膜の導電率の違いを定量的に理解することを目的とした．我々は親水部および疎水部を構成する主要な元素の数を高分子の共重合比も考慮して算出し、それらを特徴量として導電率の予測ができるか検討を行った。本研究では，解釈性に重点を置き，線形かつ低次元の解析モデルを用いて，導電性を予測するモデルの構築を試みる．一方で，どのような特徴量が導電性の予測に重要であるかは自明ではない．そこで我々は，図\ref{fig:abstract}に示すように，特徴候補を複数個用意し，導電性の予測だけでなく重要な特徴量の選択も同時に行う．

\section{Dataset outline}\label{sec:data-outline}
In this study, the target variable was defined as the hydroxide ion conductivity of anion-conductive polymer membranes measured in water at 30${}^\circ \mathrm{C}$, along with the structural characteristics of their hydrophilic and hydrophobic domains. It is well known that anion conductivity correlates with several physical parameters of the polymer membranes, including ion exchange capacity (IEC), water uptake, and the size and connectivity of hydrophilic clusters. Generally, a higher IEC leads to an increased carrier concentration and, consequently, higher conductivity. However, excessive IEC can result in increased water uptake—particularly in terms of the number of water molecules adsorbed per ionic group—thereby reducing the effective carrier density and leading to lower conductivity.

\begin{figure}
    \centering
    \includegraphics[width=\linewidth]{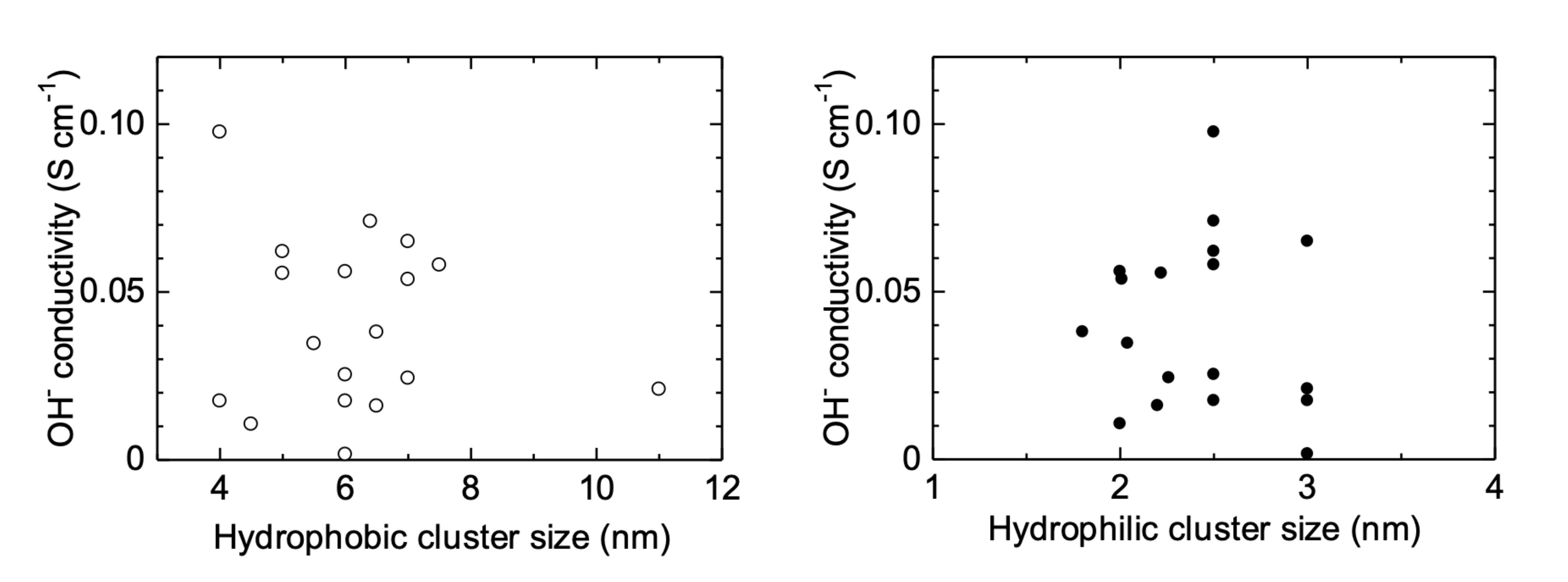}
    \caption{Scatter plot of conductivity against the average diameter of (left) hydrophobic cluster size and (right) hydrophilic cluster size obtained from TEM images \cite{MiyatakeData2018, MiyatakeData2022, MiyatakeData2023, MiyatakeData2024}.}
    \label{fig:cluster_size}
\end{figure}

The hydrophilic clusters in anion-conductive polymer membranes can be visualized from transmission electron microscopy (TEM) images of samples in which the ammonium groups were selectively stained by ion exchange with $\mathrm{PtCl_{4}^{2-}}$ ions. It is expected that larger hydrophilic clusters provide more continuous conduction pathways, leading to higher ion mobility and thus higher conductivity. Figure \ref{fig:cluster_size} shows scatter plots of hydroxide ion conductivity versus the average diameter of hydrophilic and hydrophobic clusters estimated from TEM images. However, no clear correlation was observed in either case, indicating that cluster size alone is insufficient to describe conductivity, especially when the polymer structures differ.

We hypothesize that the number of carbon (C) and fluorine (F) atoms in the hydrophilic and hydrophobic units influences the spatial occupation, domain size, and connectivity within the polymer structure, thereby functioning as key structural descriptors governing hydroxide ion conductivity. Therefore, in this study, we calculated the number of C and F atoms based on the copolymer composition for each segment and investigated whether these values could serve as predictive features for conductivity. The dataset used in this study consisted of 18 polymer samples \cite{MiyatakeData2018, MiyatakeData2022, MiyatakeData2023, MiyatakeData2024}, and the number of explanatory variables (features) was 15. We refer you to the subsection \ref{subsec:description} for details on features.

% 本研究では，アニオン導電性高分子の親水部構造、疎水部構造、およびそれらの膜の30度の水中における導電率を目的変数とする。アニオン導電率は、高分子薄膜のイオン交換容量、含水率、および親水部クラスターの大きさ・連結性、などと相関があることが知られている。一般的にイオン交換容量が大きいとキャリア濃度が大きいので導電率は高くなる。他方、イオン交換容量が過度に大きくなると含水率（特に、イオン交換基あたりの吸着水分子数）が大きくなり、実質的なキャリア密度が低下してしまうので、導電率は低くなる。

% アニオン導電性高分子薄膜の親水部クラスターは、PtCl42-イオンでアンモニウム基を染色（イオン交換した）サンプルのTEM像から求めることができる。親水クラスターが大きいとイオンが伝導する経路が十分に確保されるために移動度が大きくなるために、導電率が高くなることが期待できる。図\ref{fig:cluster_size}にTEM像から求めた親水クラスターおよび疎水クラスターの平均直径に対して導電率をプロットした。いずれにおいても明確な相関は認められず、高分子の構造が異なるとクラスター径だけでは導電率を明確に記述することができなかった。

% アニオン導電性高分子薄膜においてCやFの元素数が，高分子構造における空間的占有率や疎水・親水ドメインの大きさ・連結性に関係し，導電性を支配する構造的要因として機能すると考えられる。そこで本研究において，我々は親水部および疎水部を構成するC, Fの元素数を高分子の共重合比も考慮して算出し、それらを特徴量として導電率の予測ができるか検討を行った．本研究で用いたデータセットはサンプル数は18個であり，説明変数（特徴量）の次元は15次元であった．特徴量についてはサブセクション\ref{subsec:description}を参照してください．

\section{Method}

\subsection{Analysis outline}
\begin{figure}
    \centering
    \includegraphics[width=\linewidth]{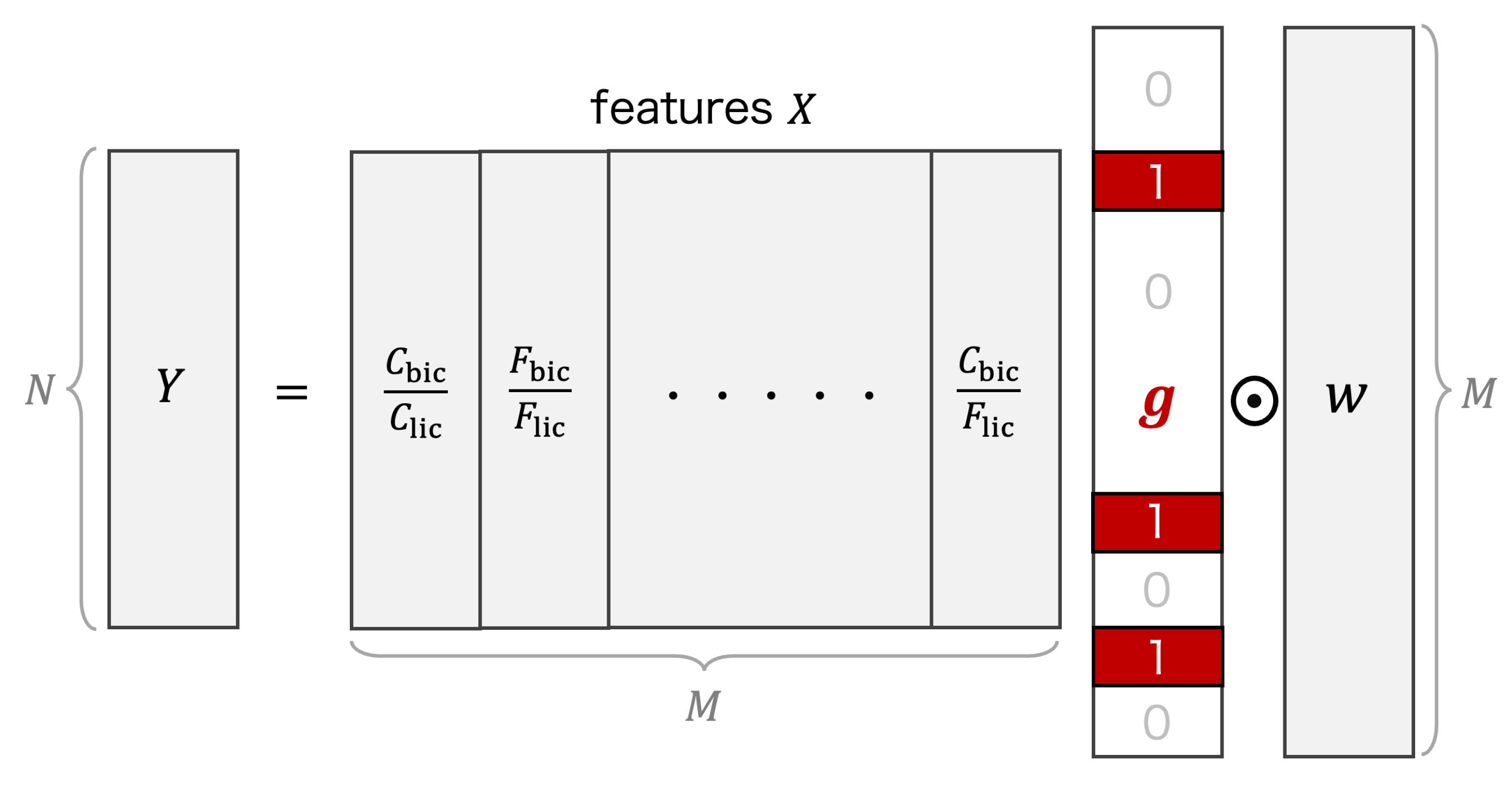}
    \caption{An overview diagram of linear regression with indicators to select a model (feature combination). $Y$ and $X$ are object and explanatory variable (candidate features), respectively. $g$ and $w$ are indicator (0 or 1) vector and weight coefficient parameters. $N$ and $M$ are the number of data samples and candidate features, respectively.}
    \label{fig:method}
\end{figure}

Our purpose is to identify explanatory features that describe the hydroxide ion conductivity of anion-conductive polymer membranes. Figure \ref{fig:method} shows an overview of the analytical workflow. As illustrated in Figure \ref{fig:method}, we focus on linear models from the perspectives of interpretability and robustness. We hypothesize that, in local regions of the design space with finite data, meaningful correlations can be captured even with a linear model, provided the constructed features contain sufficient information. Accordingly, we begin by using interpretable and controllable compositional features to predict conductivity via linear regression. The details of the regression model are described in the following subsection.

By constructing a predictive model based on compositional features that can be experimentally controlled, we aim to support the design of future experiments. However, in the materials domain, datasets often contain a limited number of samples, and including too many features can lead to overfitting and poor generalization performance. At the same time, it is not obvious which features are most strongly correlated with conductivity.

To address this, as shown in Figure \ref{fig:method}, we generate multiple candidate features and perform not only conductivity prediction but also simultaneous feature selection. For this purpose, we employ a Bayesian linear regression model with automatic feature selection. The methodology for selecting relevant features is described in the next section. By identifying a minimal set of features that can describe conductivity, we can suppress overfitting and obtain a more robust model. Furthermore, it is well known that lower-dimensional feature spaces are advantageous when designing follow-up experiments.

% 我々の目的は，アニオン導電性高分子膜における導電率を記述する特徴量を探索することである．図\ref{fig:method}に解析の概要図を示す．図\ref{fig:method}に示すように，我々は解釈性と堅牢性の観点から線形モデルに着目する．設計した特徴量に十分な情報があり，有限データが存在する局所空間では，線形でも相関を見出せるであろうという仮説のもとで解析する．まず我々は制御可能かつ解釈可能な組成情報を用いて線形回帰モデルで導電率を予測する．次節で回帰モデルについて説明する．

% 我々は制御可能な組成情報で導電率を予測することで，次の実験計画を練ることができる．しかしながら，材料分野のような少数の実験データセットにおいては，特徴量の数が多いと過学習を起こし，予測性能が低下することがしばしばある．一方で，どのような特徴量が導電率に相関するのかは自明ではない．

% そこで我々は，図\ref{fig:method}にその概要図を示すように，特徴量候補を複数個用意し，導電率の予測だけでなく重要な特徴量の選択も同時に行う．我々は特徴量選択付きのベイズ線形回帰モデルを用いた．特徴量の選択方法は，次節で説明する．必要最低数の特徴量で導電率を記述することにより，過学習を防ぐことができる．また，実験計画においても，特徴量の次元数が少ないほうが良いことが知られている．

\subsection{Description of features}\label{subsec:description}
In this study, we calculated the number of carbon (C) and fluorine (F) atoms in the hydrophilic and hydrophobic units of each polymer, taking into account the copolymer composition, and investigated whether these values could serve as effective features for predicting hydroxide ion conductivity. This approach is based on the hypothesis that the number of C and F atoms is related to the spatial occupancy of the polymer structure, as well as to the size and connectivity of hydrophilic and hydrophobic domains—factors that are believed to govern hydroxide ion conductivity.

In particular, the number of carbon atoms is expected to influence the rigidity or flexibility of the polymer backbone and side chains, the intermolecular distances, and the ability to form well-organized domain structures. These, in turn, affect the continuity and hydration of the hydrophilic network. On the other hand, the number of fluorine atoms may influence the hydrophobicity and electron density of the polymer, potentially affecting the formation and stability of ion-conducting pathways.

We examined two definitions for distinguishing hydrophilic and hydrophobic units:
\begin{itemize}
    \item *1) Direct assignment based on the chemical structure of the repeating units in the polymer backbone;
    \item *2) A more selective definition in which only side chain structures containing ammonium groups (composed exclusively of $\mathrm{sp}^3$-hybridized carbon atoms and presumed to readily form hydrophilic domains) are classified as hydrophilic, and all other structures are classified as hydrophobic.
\end{itemize}
These two definitions are hereafter referred to as *1 and *2, respectively.
% 本研究では，親水部および疎水部を構成するC, Fの元素数を高分子の共重合比も考慮して算出し、それらを特徴量として導電率の予測ができるか検討を行った。これは，CやFの元素数が，高分子構造における空間的占有率や疎水・親水ドメインの大きさ・連結性に関係し，導電性を支配する構造的要因として機能すると考えられるためである。特に炭素原子数は，主鎖や側鎖の剛直性・柔軟性，分子間距離，およびドメイン構造の形成能に影響を与え，親水性ネットワークの連続性や水和性に関係すると推察される。また，フッ素原子数は高分子の疎水性や電子密度に影響し，イオン輸送経路の形成・安定性に影響する可能性がある。

% なお、親水部と疎水部の定義としては、*1) 高分子を構成する繰り返しユニットの構造をそのまま親水部と疎水部とする、*2) sp3混成結合炭素のみから含み親水ドメインを形成しやすいことが予想されるアンモニウム基を含む側鎖構造のみを親水部とし、それ以外の構造を全て疎水部とする、という2つの定義を採用して検討を行った。それぞれの定義を記号*1および*2で示す．

\subsection{Bayesian linear regression with feature selection}\label{method:selection}
% 特徴量の選択を伴うベイズ線形回帰モデル\cite{nagata2015, igarashi2018, obinata2022}は，入力（特徴量）$x_n \in \mathbb{R}^{M}$と出力（材料特性）$y_n \in \mathbb{R}$で，以下のように表現することができる：
A Bayesian linear regression model with feature selection \cite{nagata2015, igarashi2018, obinata2022} can be formulated using input features $x_n \in \mathbb{R}^{M}$ and output material properties $y_n \in \mathbb{R}$ as follows:
\begin{equation}
    y_n = \boldsymbol{x}_n (\boldsymbol{g} \odot \boldsymbol{w})^{\top} + \varepsilon_n, \:\:\: \varepsilon_n \sim \mathcal{N}(\mu=0, \sigma^2=\lambda^{-1}).
\end{equation}
where $\boldsymbol{g} \in \{0,1\}^{M}$ is a binary indicator vector used to select relevant features, $\boldsymbol{w} \in \mathbb{R}^{M}$ is the weight coefficient vector, and $\lambda \in \mathbb{R}$ is the precision parameter. The operator $\odot$ denotes the Hadamard (element-wise) product. When $g_m = 1$, the $m$-th feature $x_{:, m}$ is considered necessary for predicting the material property; conversely, $g_m = 0$ indicates that the feature is not required. The noise term $\varepsilon_n$ is assumed to follow a normal distribution, $\mathcal{N}(\mu=0, \sigma^2 = \lambda^{-1})$. Accordingly, the likelihood of the dataset $\mathcal{D}=\{(\boldsymbol{x}_n, y_n)\}_{n=1}^{N}$ consisting of $N$ samples can be expressed as:
% ここで，$\boldsymbol{g} \in \{0,1\}^{M}$は特徴量を選択するインジケータ，$\boldsymbol{w} \in \mathbb{R}^{M}$は重み係数パラメータ，$\lambda \in \mathbb{R}$は精度パラメータ，演算子$\odot$はアルマーダ積である．$g_m=1$の場合は，特徴量$x_{:, m}$が材料特性の予測に必要であることを意味する．逆に，$g_m=0$の場合は，不必要であることを意味する．$\varepsilon_n$は統計ノイズであり，正規分布$\mathcal{N}(\mu=0, \sigma^2=\lambda^{-1})$に従うと仮定する．従って，$N$個から成る出力データ$\mathcal{D}=\{(\boldsymbol{x}_n, y_n)\}_{n=1}^{N}$の確率分布を以下のように表現できる：
\begin{equation}
    P(\mathcal{D} \mid \boldsymbol{g}, \boldsymbol{w}, \lambda) = \prod_{n=1}^{N}{P(y_n \mid  \boldsymbol{g}, \boldsymbol{w}, \lambda, \boldsymbol{x}_n)} = \prod_{n=1}^{N}{\mathcal{N}(\mu = \boldsymbol{x}_n (\boldsymbol{g} \odot \boldsymbol{w})^{\top}, \sigma^2 = \lambda^{-1})}.
\end{equation}
% したがって，インジケータ$\boldsymbol{g}$の事後分布$P(\boldsymbol{g} \mid Y)$は，下記のように表現される：
Accordingly, the posterior distribution of the indicator vector $\boldsymbol{g}$, denoted as $P(\boldsymbol{g} \mid Y)$, can be expressed as follows:
\begin{equation}\label{eq:Marginal_loglikelihood}
    P(\boldsymbol{g} \mid \mathcal{D}) \varpropto P(\mathcal{D} \mid \boldsymbol{g})P(\boldsymbol{g}) = \iint{P(\mathcal{D} \mid  \boldsymbol{g}, \boldsymbol{w}, \lambda)P(\boldsymbol{g})P(\boldsymbol{w})P(\lambda)}\mathrm{d}\boldsymbol{w}\mathrm{d}\lambda
\end{equation}
% ここで，$P(g)$, $P(w)$, $P(\lambda)$は各パラメータの事前分布である．事前分布は下記のように設定した．
Here, $P(\boldsymbol{g})$, $P(\boldsymbol{w})$, and $P(\lambda)$ represent the prior distributions for each parameter. These priors are defined as follows:
\begin{align}
    P(\boldsymbol{g}) &= \prod_{m=1}^{M}{\mathcal{B}(g_m \mid p_{\mathrm{B}})},\\
    P(\boldsymbol{w}) &= \prod_{m=1}^{M}{\mathcal{N}(w_m \mid \mu_{\mathrm{N}}, \sigma^2_{\mathrm{N}})},\\
    P(\lambda) &= \mathcal{G}(\lambda \mid k_{\mathrm{G}}, \theta_{\mathrm{G}}),
\end{align}
where $\mathcal{B}(g \mid p_{\mathrm{B}})$, $\mathcal{N}(w \mid \mu_{\mathrm{N}}, \sigma^2_{\mathrm{N}})$, and $\mathcal{G}(\lambda \mid k_{\mathrm{G}}, \theta_{\mathrm{G}})$ represent the Bernoulli, Normal, and Gamma distributions, respectively. In equation (\ref{eq:Marginal_loglikelihood}), marginalization over the weight parameter $\boldsymbol{w}$ was performed analytically, while marginalization over the precision parameter $\lambda$ was carried out using numerical integration.
% ここで，$\mathcal{B}(g \mid p_{\mathrm{B}})$と$\mathcal{N}(w \mid \mu_{\mathrm{N}}, \sigma^2_{\mathrm{N}})$，$ \mathcal{G}(\lambda \mid k_{\mathrm{G}}, \theta_{\mathrm{G}})$はそれぞれベルヌーイ分布と正規分布，ガンマ分布である．式(\ref{eq:Marginal_loglikelihood})において，重み係数パラメータ$w$に関する周辺化（積分操作）は解析的に計算を行った．また，精度パラメータ$\lambda$に関する周辺化（積分操作）は数値積分により行なった．

% 本研究では，式(\ref{eq:Marginal_loglikelihood})の確率に従い交換モンテカルロ法\cite{Geyer1991,Swendsen1986,Hukushima1996}でインジケータを近似的に全状態探索する．ここで，$m$番目の特徴量に着目したインジケータの確率は以下のようにで表現される．：
In this study, we approximated an exhaustive search over all possible indicator states using the exchange Monte Carlo method \cite{Geyer1991,Swendsen1986,Hukushima1996}, based on the probability defined in equation (\ref{eq:Marginal_loglikelihood}). The marginal posterior probability that the $m$-th feature is selected can be expressed as follows:
\begin{equation}
    P(g_m=1 \mid \mathcal{D}) = \sum_{\{\boldsymbol{g} \mid g_m=1\}}{P(\boldsymbol{g} \mid \mathcal{D})}
\end{equation}

\section{Experimental configuration}
In this study, we used the exchange Monte Carlo method to estimate the post-distribution of indicators. The burn-in period was set to 10000 iterations, and 10000 samples were drawn for inference. In the exchange monte carlo method, we set the reverse temperature $\beta$ as follows: $\beta = \{\rho^{t-T} \mid t=\{1, 2, \cdots, T\} \}$ where the number of temperature $T$ and the proportion $\rho$ were set to $T=46$ and $\rho=1.1$, respectively.

We set the hyperparameters of the prior distribution as follows: $p_{\mathrm{B}}=0.5$, $\mu_{\mathrm{N}}=0$, $\sigma^2_{\mathrm{N}}=1.0$, $k_{\mathrm{G}}=1.5$, $\theta_{\mathrm{G}}=10$. The computations were performed on a MacBook Air equipped with an Apple M2 chip and 24 GB of RAM.

\begin{figure}
    \centering
    \includegraphics[width=0.8\linewidth]{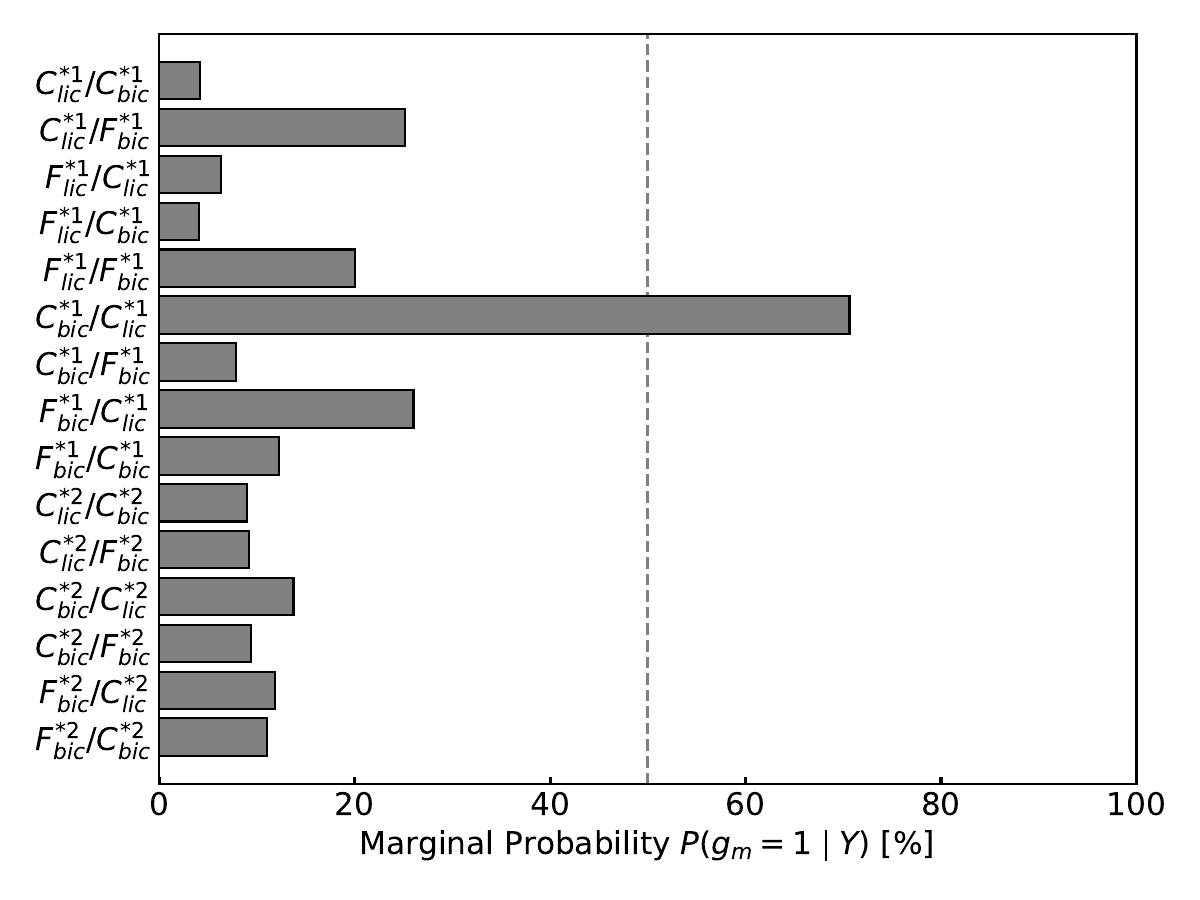}
    \caption{Result of feature selection based on the marginal log-likelihood. Candidate features was defined from composition information. The wavy line indicates the prior probability of 50\%. Object variable is conductivity.}
    \label{fig:selection}
\end{figure}

\begin{figure}
    \centering
    \includegraphics[width=0.8\linewidth]{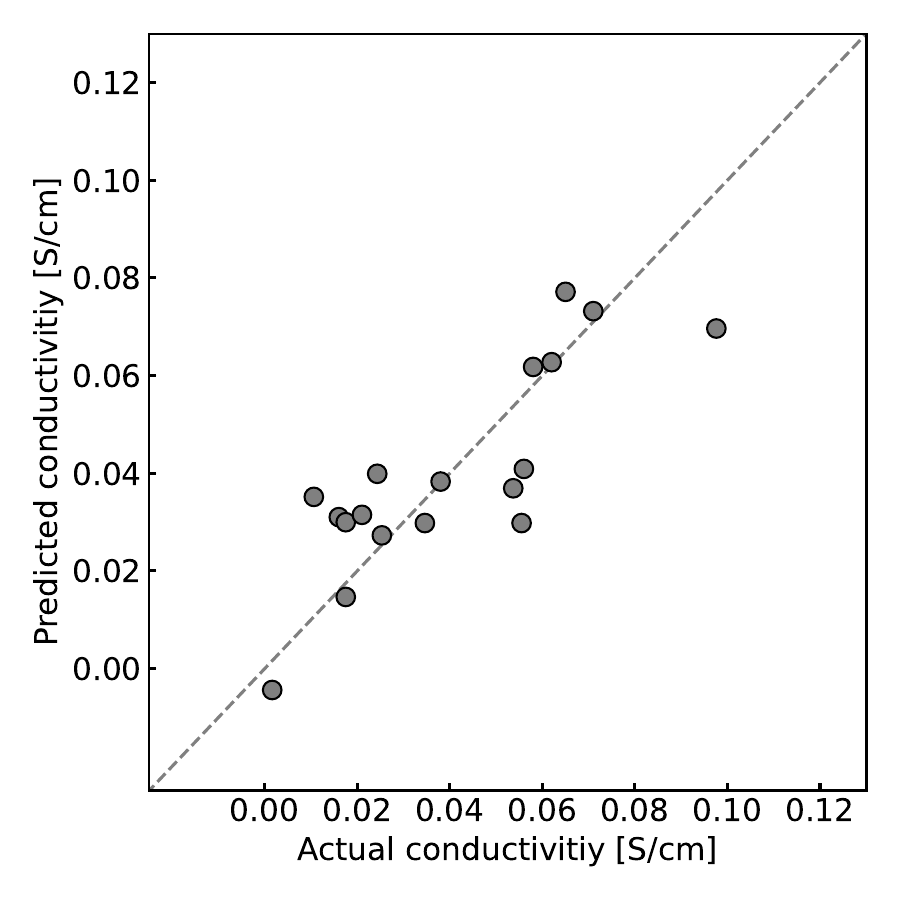}
    \caption{Conductivity prediction result based on selected model: $\hat{y} = a C_{bic}^{(*1)}/C_{lic}^{(*1)}+b$ where $a \approx -0.035$ and $b \approx 0.09$.}
    \label{fig:prediction}
\end{figure}

\section{Results and discussion}
The results of feature selection are shown in Figure \ref{fig:selection}. In this figure, the y-axis represents the labels of candidate features, and the x-axis corresponds to the marginal posterior probability of the indicator variable for each feature, denoted as $P(g_m = 1 \mid \mathcal{D})$. Since the prior distribution for the indicator variable was set as a Bernoulli distribution $\mathcal{B}(p_{\mathrm{B}}=0.5)$, a posterior probability exceeding 50\% indicates that the corresponding feature is considered important for the prediction task.

From Figure \ref{fig:selection}, it is evident that the ratio of carbon atoms in the hydrophobic and hydrophilic units, $C_{bic}^{(*1)}/C_{lic}^{(*1)}$, is identified as the most significant feature for predicting hydroxide ion conductivity in this dataset. The marginal probabilities for all other features remain below 50\%, suggesting that $C_{bic}^{(*1)}/C_{lic}^{(*1)}$ alone captures the underlying trend of conductivity in this dataset. This result is non-trivial and highlights the unique importance of this compositional ratio.

Accordingly, the trend of hydroxide ion conductivity in this dataset can be expressed as follows:
\begin{equation}\label{eq:result}
    \hat{y} = a \frac{C_{bic}^{(*1)}}{C_{lic}^{(*1)}}+b,
\end{equation}
where the estimated coefficients were $a \approx -0.035$ and $b \approx 0.09$. This finding—that the ratio of carbon atoms in the hydrophobic and hydrophilic units, $C_{bic}^{(*1)}/C_{lic}^{(*1)}$, exhibits a strong correlation with hydroxide ion conductivity—is novel. Within the present dataset, the carbon content ratio was found to be the most influential factor in determining conductivity, compared to other elemental compositions. This result is also supported by physicochemical reasoning.

It would be probably reasonable that when the hydrophobic units contain a large number of carbon atoms, the hydrophobic domains within the polymer membrane become more dominant. This expansion reduces the connectivity and size of the hydrophilic domains, which can interrupt continuous hydroxide ion transport pathways and consequently lower ionic mobility. In contrast, when the hydrophilic units contain a relatively greater number of carbon atoms, the hydrophilic domains—particularly those containing ammonium groups—tend to become larger, more hydrated, and better connected, thereby enhancing ion transport efficiency. Therefore, the ratio $C_{bic}^{(*1)}/C_{lic}^{(*1)}$ can be interpreted as a descriptor for the formation of effective ion transport pathways, and it exerts a strong influence on hydroxide ion conductivity as a result.

From Figure \ref{fig:selection}, it was found that, between the following two definitions of hydrophilic and hydrophobic units. The *1 definition yielded more effective features for predicting hydroxide ion conductivity, as indicated by the data-driven analysis. Figure \ref{fig:prediction} shows the regression results of the predictive model constructed based on the marginal posterior probability. As shown in Figure \ref{fig:prediction}, the model, despite being based on a single feature, is capable of capturing the trend in conductivity. The RMSE (Root Mean Squared Error) was 0.014 [S/cm].

We successfully constructed an interpretable predictive model for this dataset. In materials informatics, where the number of available data points is often limited, predictive models with a small number of features are advantageous because they reduce the risk of overfitting and provide more robust predictions. Therefore, such simplified models can serve as valuable tools for guiding subsequent experimental design.

It should be noted that the dataset used in this study consists of only 18 samples, and the resulting predictive model reflects the trends observed within this specific dataset. Therefore, to evaluate the generalizability and reproducibility of the model, further accumulation of experimental data and validation using external datasets are required.

\begin{figure}
    \centering
    \includegraphics[width=0.8\linewidth]{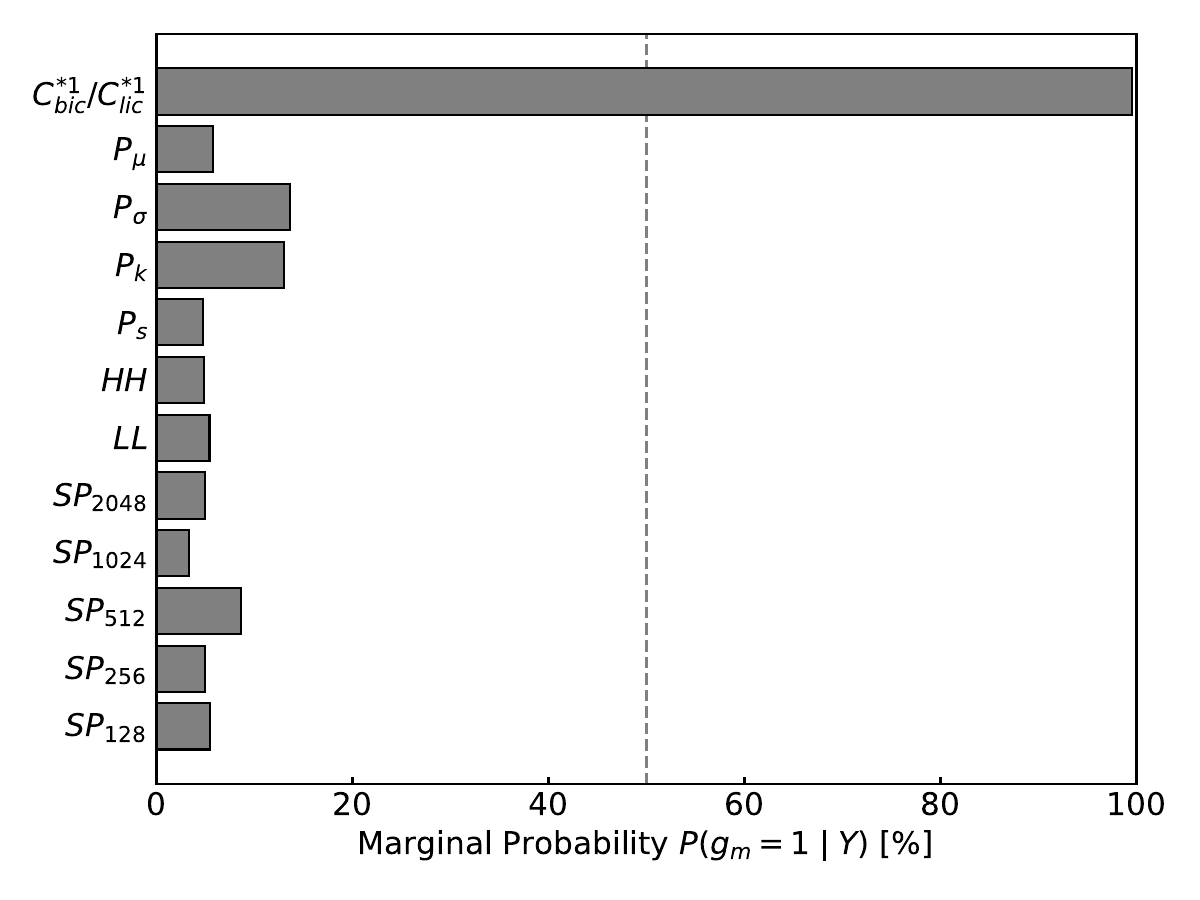}
    \caption{Result of feature selection based on the marginal log-likelihood. Candidate features was $C_{bic}^{(*1)}/C_{lic}^{(*1)}$ and frequency features of TEM images. The wavy line indicates the prior probability of 50\%. Object variable is conductivity. We used the TEM image features as follows: brightness statistics ${P_{\mu}, P_{\sigma}, P_k, P_s}$, and frequency-domain features ${HH, LL, SP_{2048}, SP_{1024}, SP_{256}, SP_{128}}$ obtained via wavelet transformation \cite{Texture2021}. The $SP_{*}$ is an abbreviation for the steerable pyramids, and the subscript indicates the size of the filtered image.}
    \label{fig:ae1:selection}
\end{figure}

\section{Additional Experiments}\label{sec:tem}
\subsection{Attempt to build a prediction model with added TEM features}
In this study, we successfully constructed an interpretable predictive model based on the present dataset. However, as shown in Figure \ref{fig:prediction}, there are several data points around 0.04 [S/cm] where the model exhibits low sensitivity. This observation suggests that while the primary trend in conductivity can be captured by the compositional feature $C_{bic}^{(*1)}/C_{lic}^{(*1)}$, there may be secondary trends that are not fully accounted for. To address this, we attempted to extract complementary information from transmission electron microscopy (TEM) images that could augment the compositional descriptor $C_{bic}^{(*1)}/C_{lic}^{(*1)}$.

TEM images are high-dimensional data, as each pixel represents a feature. Given that the TEM images of the anion-conductive membranes exhibit patterned textures, we extracted two types of features: brightness statistics ${P_{\mu}, P_{\sigma}, P_k, P_s}$, and frequency-domain features ${HH, LL, SP_{2048}, SP_{1024}, SP_{256}, SP_{128}}$ obtained via wavelet transformation \cite{Texture2021}. Each feature is explained in Appendix \ref{appx:img_feature}.

To extract the frequency features $\mathcal{F}$, we employed the pyrtools library \cite{pyrtools, Simoncelli1995}, a powerful toolkit for multiscale image analysis. Specifically, Gabor filters with varying scales and orientations were applied, and the spatial average of each filtered image was used as a feature. The number of scales and orientations was set to 5 and 4, respectively. Since the TEM images of the membranes do not exhibit strong orientation preferences, we averaged the results across all orientations. In this analysis, we defined the explanatory variables as $X = {C_{bic}^{(*1)}/C_{lic}^{(*1)}} \cup \mathcal{F}$, and conducted Bayesian linear regression with automatic feature selection.

Figure \ref{fig:ae1:selection} shows the results of feature selection. The y-axis represents feature labels, and the x-axis indicates the marginal posterior probability for each feature, $P(g_m = 1 \mid \mathcal{D})$. As seen in Figure \ref{fig:ae1:selection}, the compositional feature $C_{bic}^{(*1)}/C_{lic}^{(*1)}$ was selected with nearly 100\% probability, whereas all frequency-domain features from set $\mathcal{F}$ were selected with probabilities below 50\%.

These results suggest that the compositional feature already encapsulates the essential structural information that would otherwise be captured from the TEM images. In other words, the frequency features derived from the images do not provide additional information for predicting hydroxide ion conductivity. This implies that the banded structures observable in the TEM images—reflected in the frequency-domain features—are predominantly determined by the underlying chemical structure. The verification of this suggestion is shown in Appendix \ref{appx:corr}.

% 我々は，本データセットにおいて，解釈性に優れた予測モデルを構築することに成功した．しかしながら，図\ref{fig:prediction}において，0.04[S/cm]付近では，予測モデルの感度が小さいサンプルデータがいくつかあることが確認できる．このことから，導電性の主要な傾向は，組成情報$C_{bic}^{(*1)}/C_{lic}^{(*1)}$で記述できるが，副次的な傾向は他にあると推察される．そこで本研究では，TEM画像に着目して組成情報$C_{bic}^{(*1)}/C_{lic}^{(*1)}$の相補的な情報の抽出を試みた．TEM画像はピクセル数の特徴量次元を持つため，高次元なデータある．そこで本研究では，アニオン性膜のTEM画像がパターン画像であることから，輝度統計量$\{P_{\mu}, P_{\sigma}, P_k, P_s\}$およびTEM画像からウェーブレット変換を用いて周波数特徴量$\{HH, LL, SP_{2048}, SP_{1024}, SP_{256}, SP_{128}\}$を抽出する．

% 本研究では，画像の多重解像度解析の有力なライブラリであるpyrtoolsを用いて周波数特徴量セット$\mathcal{F}$を抽出する．具体的には，異なるスケールや方位を持つガボールフィルタを適用して，得られたフィルタ画像の空間平均を特徴量とする．スケールと方位の個数はそれぞれ5と4に設定した．アニオン性膜のTEM画像は方位選択制がないため，方位に対して平均処理を行った．本章では，説明変数$X=\{C_{bic}^{(*1)}/C_{lic}^{(*1)}\} \cup \mathcal{F}$として，特徴量選択付きベイズ線形を行なった．図\ref{fig:ae1:selection}に特徴量の選択結果を示す．

% 図\ref{fig:ae1:selection}のy軸は特徴量のラベルであり，x軸は１つの特徴量に着目したインジケータの周辺確率$P(g_m=1 \mid \mathcal{D})$である．図\ref{fig:ae1:selection}から，組成情報$C_{bic}^{(*1)}/C_{lic}^{(*1)}$はほぼ100\%の確率で選ばれており，周波数特徴量セットFの確率がすべて50\%以下であることがわかる．周波数特徴量の周辺確率が低いことから，本研究で用いた組成情報がTEM画像の主要な情報をすでに内包しており、周波数特徴量は導電率予測に対して追加的な情報を与えていないことが示唆された。これは，導電性に寄与するTEM画像のバンド構造（周波数特徴量）は主に化学構造により決定されることを示唆している．

\section{Conclusion}
This study aimed to identify key features that describe hydroxide ion conductivity in anion-conductive polymer membranes. To achieve this, we constructed a predictive model using Bayesian linear regression with automatic feature selection. As a result, we found that the ratio of carbon atoms in the hydrophobic and hydrophilic units, $C_{bic}^{(*1)}/C_{lic}^{(*1)}$, was the most significant feature for predicting conductivity within the given dataset.

The linear model with feature selection developed in this work offers a promising approach for material design support in advanced polymer development, where data availability is often limited. It provides both interpretability and reusability, which are critical for practical applications. Future work will focus on validating and generalizing this approach by incorporating a broader range of polymer structures and external datasets.

% 本研究は，アニオン性高分子膜における導電率を記述する特徴量を探索することを目的とした．我々は特徴量選択付きのベイズ線形回帰を用いた予測モデルの構築を行なった．その結果，本データセットにおいて，疎水部と親水部におけるカーボン数の比率$C_{bic}^{(*1)}/C_{lic}^{(*1)}$が導電性を予測するために重要な特徴量であることを明らかにした．本研究で構築した特徴量選択付き線形予測モデルは，データ数が限られる先端材料開発の現場において，解釈性と再利用性を兼ね備えた設計支援手法として有望である。今後は，より多様な高分子構造および外部データを取り入れた検証により，このアプローチの一般化と拡張が期待される。

\section*{Acknowledgements}
This work was supported by the MEXT Program: Data Creation and Utilization-Type Material Research and Development Project (Grant Number JPMXP1122715503). The authors also acknowledge the use of the RDE system provided through the DICE service (https://dice.nims.go.jp/) of the Materials Data Platform (MDPF), National Institute for Materials Science (NIMS), Japan. We thank Ms. Hiroko Nagao (NIMS) for the implementation of the tool used in this study, which is publicly available through the RDE system as template.

\bibliographystyle{tfnlm} % 参考文献
\bibliography{main} %

\clearpage

\section{Appendix}

\subsection{Explanation of image features}\label{appx:img_feature}
This section outlines the image features extracted from the TEM data used in this study. Specifically, we derived two types of features from the TEM images; brightness statistics $P_{\mu}$, $P_{\sigma}$, $P_k$, $P_s$ and frequency-domain features $HH$, $LL$, $SP_{2048}$, $SP_{1024}$, $SP_{256}$, $SP_{128}$ obtained via wavelet transformation. A detailed description of each feature is provided in Table \ref{table:des_feat}. In this study, the pixel size of the original TEM image input was 2048x2048. In Table \ref{table:des_feat}, the $SP_{*}$ is an abbreviation for the steerable pyramids, and the subscript indicates the size of the filtered image.
% この節では，本研究で用いたTEM画像の特徴量について示す．本研究で，我々はTEM画像から輝度統計量$\{P_{\mu}, P_{\sigma}, P_k, P_s\}$およびTEM画像からウェーブレット変換を用いて周波数特徴量$\{HH, LL, SP_{2048}, SP_{1024}, SP_{256}, SP_{128}\}$を特徴量として抽出した．それぞれの特徴量の説明を表\ref{appx:des_feat}に示す．

\begin{table}
    \centering
    \caption{ Explanation of image features in this study. }
    \label{table:des_feat}
    \scalebox{1.0}[1.0]{
    \begin{tabular}{ccc}
    \hline
    Type & Name & Explanation \\
    \hline
    \multirow{4}{*}{Brightness} 
    & $P_{\mu}$    & mean of brightness distribution \\
    & $P_{\sigma}$ & variance of brightness distribution \\
    & $P_k$        & kurtosis of brightness distribution \\
    & $P_s$        & skewness of brightness distribution \\
    \hline
    \multirow{6}{*}{Frequency}
    & $HH$         & residual highpass feature \\
    & $LL$         & residual lowpass feature \\
    & $SP_{2048}$  & average magnitude of decomposition image (2048x2048) \\
    & $SP_{1024}$  & average magnitude of decomposition image (1024x1024) \\
    & $SP_{256}$   & average magnitude of decomposition image (256x256) \\
    & $SP_{128}$   & average magnitude of decomposition image (128x128) \\
    \hline
    \end{tabular}
    }
    % \begin{tablenotes}
    %   \footnotesize
    %   \item \hfill Note: BCC and FCC refer to body-centered cubic and face-centered cubic crystal structures, respectively.
    % \end{tablenotes}
\end{table}

\begin{figure}
    \centering
    \includegraphics[width=0.8\linewidth]{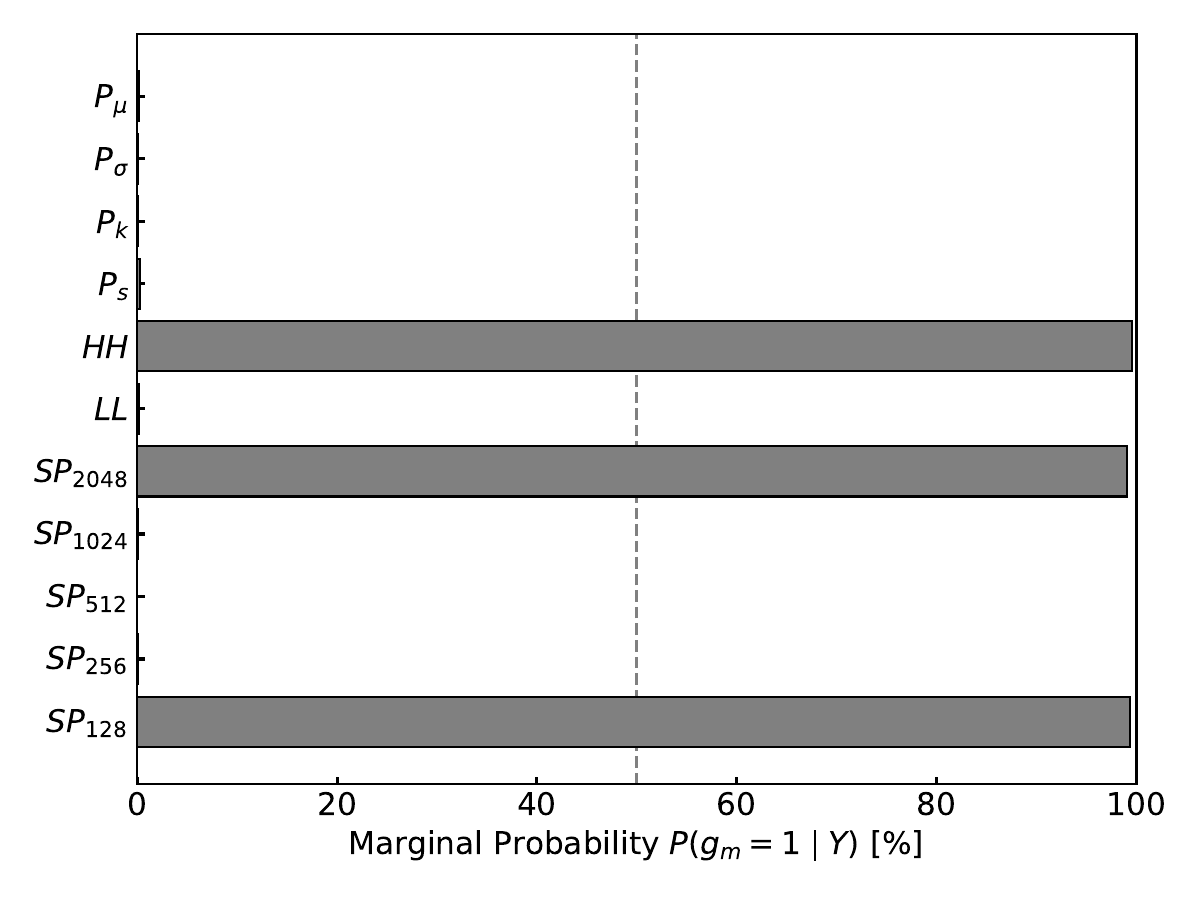}
    \caption{Result of feature selection based on the marginal log-likelihood. Candidate features was frequency features of TEM images. The wavy line indicates the prior probability of 50\%. Object variable is $C_{bic}^{(*1)}/C_{lic}^{(*1)}$.}
    \label{fig:ae2:selection}
\end{figure}

\begin{figure}
    \centering
    \includegraphics[width=0.8\linewidth]{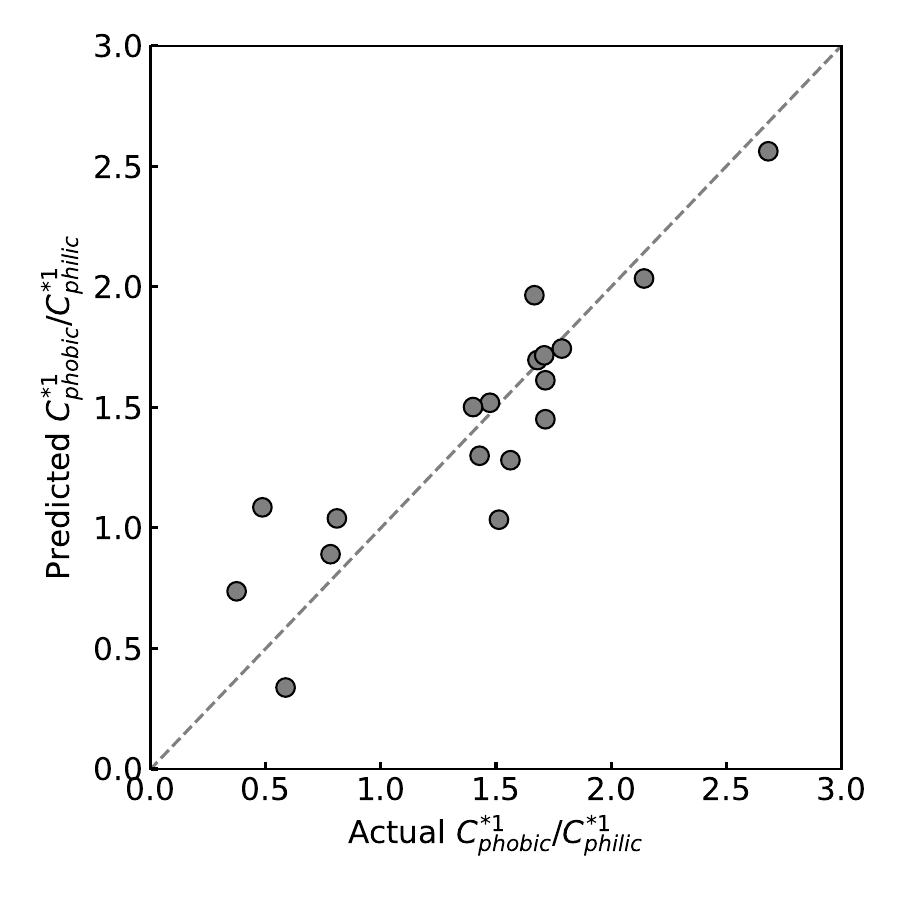}
    \caption{$C_{bic}^{(*1)}/C_{lic}^{(*1)}$ prediction result based on selected model}
    \label{fig:ae2:prediction}
\end{figure}

\begin{figure}
    \centering
    \includegraphics[width=0.8\linewidth]{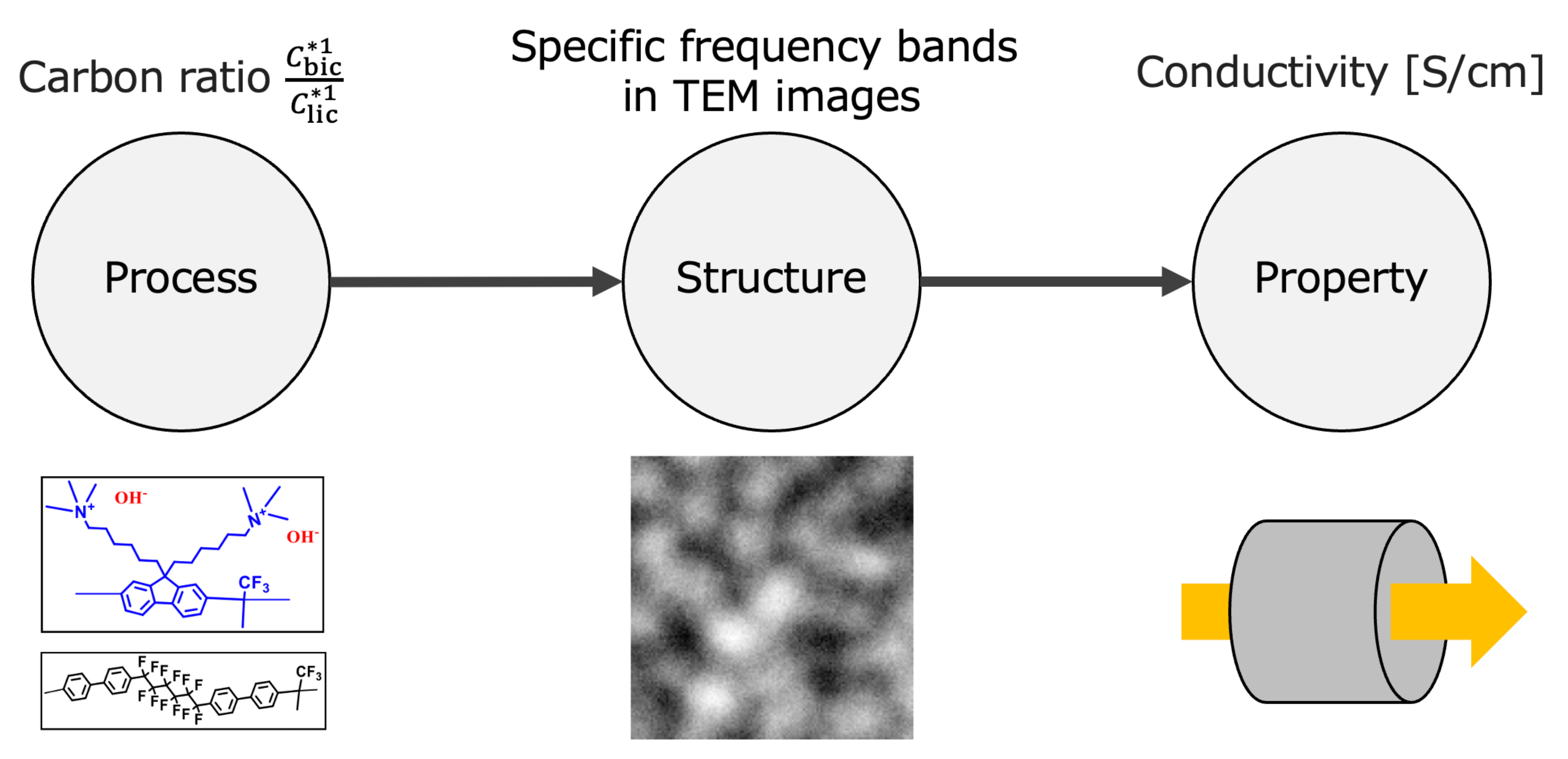}
    \caption{Relationships between processes, structures, and properties in this dataset organized based on model selection results.}
    \label{fig:ae2:sorting}
\end{figure}

\subsection{Correlation analysis between composition and TEM image}\label{appx:corr}
Section \ref{sec:tem} suggested that frequency features derived from TEM images contain complementary information which correspond to the ratio of carbon numbers in hydrophilic and hydrophobic domains. To confirm this, we constructed a predictive model in which the TEM frequency features were used as explanatory variables, and the compositional feature $C_{bic}^{(*1)}/C_{lic}^{(*1)}$ was set as the response variable. Bayesian linear regression with automatic feature selection was employed. The results of feature selection are shown in Figure \ref{fig:ae2:selection}, where the y-axis represents the feature labels and the x-axis indicates the marginal posterior probability $P(g_m = 1 \mid \mathcal{D})$ for each feature.

As shown in Figure \ref{fig:ae2:selection}, three frequency features—$HH$, $SP_{2048}$, $SP_{128}$—were selected. The corresponding regression weights $w_d$ for these features were $\{3.43, -2.68, 0.80\}$, respectively. Figure \ref{fig:ae2:prediction} presents the regression result based on the selected model estimated from the marginal probabilities.

From Figure \ref{fig:ae2:prediction}, it can be seen that the compositional feature $C_{bic}^{(*1)}/C_{lic}^{(*1)}$ can be effectively approximated using only three frequency-domain features—i.e., features corresponding to specific spatial frequency bands in the TEM images. This indicates that the frequency-domain features do not provide complementary information with respect to conductivity prediction, but rather encode similar structural information already captured by the compositional descriptor.

Structurally, when the hydrophilic domain contains a higher number of carbon atoms, the side chains tend to be more flexible and spatially extended, facilitating the formation of larger and more interconnected hydrophilic domains. In contrast, a higher carbon content in the hydrophobic domain—particularly rigid structures such as aromatic rings in the main chain—leads to an increased occupation of hydrophobic regions, which can fragment the hydrophilic domains. These structural differences are likely to manifest in the TEM images as variations in specific spatial frequency components, corresponding to domain size and connectivity.

Based on these computational experiments, the relationship among processing, structure, and properties can be summarized as shown in Figure \ref{fig:ae2:sorting}. This process–structure–property correlation analysis suggests that the compositional ratio $C_{bic}^{(*1)}/C_{lic}^{(*1)}$ controls specific spatial frequency components in TEM images, thereby enhancing hydroxide ion conductivity.

% この節では，組成情報$C_{bic}^{(*1)}/C_{lic}^{(*1)}$とTEM画像の周波数特徴量が相補的な情報を持つことを確認するために，TEM画像の周波数特徴量を用いて組成情報$C_{bic}^{(*1)}/C_{lic}^{(*1)}$を予測するモデルの構築を試みる．TEM画像の周波数特徴量を説明変数，組成情報$C_{bic}^{(*1)}/C_{lic}^{(*1)}$を目的変数として，特徴量選択付きのベイズ線形回帰を実施した．図\ref{fig:ae2:selection}に特徴量の選択結果を示す．図\ref{fig:ae2:selection}のy軸は特徴量のラベルであり，x軸は１つの特徴量に着目したインジケータの周辺確率$P(g_m=1 \mid \mathcal{D})$である．図\ref{fig:ae2:selection}から，3つの周波数特徴量$\{HH,SP_{2024},SP_{128}\}$が選ばれていることが確認できる．3つの周波数特徴量$\{HH,SP_{2024},SP_{128}\}$の重み係数$w_d$は，それぞれ$\{3.43,-2.68,0.80\}$であった．図\ref{fig:ae2:prediction}に周辺確率$P(g_m=1 \mid \mathcal{D})$で推定された予測モデルによる回帰結果を示す．

% 図\ref{fig:ae2:prediction}より，3つの周波数特徴量，つまり特定のバンド周波数の特徴量で組成情報$C_{bic}^{(*1)}/C_{lic}^{(*1)}$を表現できていることがわかる．したがって，この結果は，TEM画像の周波数特徴量は，導電性に対して，組成情報$C_{bic}^{(*1)}/C_{lic}^{(*1)}$と相補的な情報はなく，同様の情報をコードしていることが確認できた．親水部に多くのCを含む場合、側鎖の柔軟性や空間的自由度が増し、より大きく連結性の高い親水ドメインが形成されやすくなります。一方、疎水部に多くのC（主鎖、特に芳香環などの硬直構造）が含まれると、疎水ドメインの占有率が高くなり、親水部が断片化しやすくなります。このような構造的な違いは、TEM画像において特定の空間周波数（つまり，ドメインサイズや連結性に対応する構造）として現れることが考えられる．ここまでの計算実験に基づいて，プロセス-構造-特性の観点で整理すると，図\ref{fig:ae2:sorting}のように示すことができる．プロセス-構造-特性の観点から整理した相関分析を行うことで，$C_{bic}^{(*1)}/C_{lic}^{(*1)}$がTEM画像の特定バンド周波数を制御して導電性を向上させることを示唆した．

\end{document}